\documentstyle[aps,prb,epsfig,multicol]{revtex}
\begin{document}
\title{Field-Dependent Critical Current in Type-II Superconducting Strips:
Combined Effect of Bulk Pinning and Geometrical Edge Barrier}
\author{Andrey A. Elistratov, Denis Yu. Vodolazov,  Igor L. Maksimov and John
R. Clem$^*$}
\address{Nizhny Novgorod University,
  603600 Nizhny Novgorod, Russia}
\address{$^*$Ames Laboratory and Department of Physics and Astronomy, Iowa
State University, Ames, Iowa 50011--3160, USA}
\date{\today}
\maketitle

\begin{abstract} Recent theoretical and experimental research on
low-bulk-pinning superconducting strips has revealed striking dome-like
magnetic-field distributions due to geometrical edge barriers.  The observed
magnetic-flux profiles differ strongly from those in strips in which bulk
 pinning is dominant.  In this paper we theoretically describe the current
and field distributions of a superconducting strip under the combined
influence of both a geometrical edge barrier and bulk pinning at the strip's
critical current $I_c$, where a longitudinal voltage first appears.  We
calculate
$I_c$ and find its dependence upon a perpendicular applied magnetic field
$H_a$. The behavior is governed by a parameter
$p$, defined as the ratio of the bulk-pinning critical current $I_p$ to the
geometrical-barrier critical current $I_{s0}$.  We find that when $p > 2/\pi$
and $I_p$ is field-independent,
$I_c$ vs $H_a$ exhibits a plateau  for small $H_a$, followed by the dependence
$I_c-I_p
\propto H_a^{-1}$ in higher magnetic fields.
\end{abstract}
\begin{multicols}{2}
\narrowtext

The combination of a
geometrical edge barrier and bulk pinning recently  has been shown to strongly
affect the properties of low-dimensional superconductors (thin films, single
crystals, and tapes with high demagnetizing factors) placed in either a
perpendicular magnetic field
\cite {Zeldov+,MakELisJLet,Benkraouda96,MakELisAPL1,ElisMaks} or a
transport-current-carrying state
\cite {Kupr,ILMakEPL,Benkr,MakELisAPL2}. 
While most experimental studies of 
the field dependence of the critical current
$I_c$ are being interpreted solely on the basis of bulk-pinning theory (see
for example
\cite{Weyers,Muller,Aubin,Brion,Lefloch}), a number of works
\cite{Kupr,Benkr,Andr,Burlach} have shown that a geometrical edge barrier (or
surface barrier) may strongly affect 
$I_c$. In this paper we
study the combined effect of a geometrical  edge barrier and bulk pinning upon
the magnetic field dependence of $I_c$ for  type-II superconducting strips. We
shall show  how the dependence of $I_c$ upon $H_a$ is controlled by the
parameter
$p=I_p/I_{s0}$, where $I_p $ is the  bulk-pinning critical current in the
absence of a geometrical edge barrier, and $I_{s0}$ is the geometrical-barrier
critical current in the absence of bulk pinning.

We consider a superconducting strip of thickness $d$ ($|y| < d/2$) and
width $2W$ ($ |x| < W$) centered on the
$z$ axis.  We assume that $d$ is less than the London penetration depth
$\lambda$ and that $W$ is much larger than the two-dimensional screening
length $\Lambda = 2 \lambda^2/d$.   The strip is subjected to a perpendicular
applied magnetic field
${\bf H}_a = (0,H_a,0)$, and it carries  a total current
$I$ in the $z$ direction described by a spatially dependent sheet current
density
${\bf K}(x) = {\bf J}d = [0,0,K_z(x)]$. We wish to determine the
current-density and magnetic-field distributions at the critical current at
which a steady-state flux-flow voltage appears along the length of the strip.  
For a strip containing no magnetic flux,
$K_z(x)$ is the sum of two contributions,\cite{Benkr}
\begin{equation} K_{Iz}(x)=\frac{I}{\pi\sqrt{W^2-x^2}},
\label{iI}
\end{equation} the Meissner-state current density generated by the applied
current $I$, and
\begin{equation} K_{az}(x)=\frac{2H_ax}{\sqrt{W^2-x^2}},
\label{iH}
\end{equation} the Meissner-state current density induced by the applied field
$H_a$. The divergences in Eqs.\ (\ref{iI}) and (\ref{iH}) at $|x| = W$ are cut
off when
$x$ is within $\Lambda$ of the edge.

To account  for  the edge barrier, we assume that vortices
 nucleate and enter the superconductor when $K_z$ at either sample edge
reaches the value $K_s = j_s d$ at which the barrier is overcome.  For an
ideal edge,  $j_s$ is equal to the Ginzburg-Landau depairing current density
$j_{GL}$ 
\cite{Aslamazov,VMB}, but for an extremely defected edge, $j_s$ may become
negligibly small.  When
$H_a=0$, the sheet current at both edges is approximately $K_{Iz} \approx I/
\pi
\sqrt{2W\Lambda}$, such that the edge-barrier critical current in zero
external magnetic field is
$I_{s0} \approx
\pi K_s \sqrt{2W \Lambda}$. When $H_a>0$, the net sheet current at $x = W$ is
approximately 
$K_z \approx (I+2 \pi H_a W)/ (\pi \sqrt{2W \Lambda})$, and the edge-barrier
critical current becomes
$I_s(H_a)/I_{s0} = 1-h$ for small $h$, where $h = H_a/(I_{s0}/2 \pi W).$  This
result is essentially the same as that found in Ref.\cite{Benkr} for the
critical current in low applied magnetic fields for bulk-pinning-free strips.

We next account for bulk pinning, characterized via a bulk-pinning critical
sheet current density, $K_p = j_p d$, such that the critical current in the
absence of an edge barrier is $I_p = 2K_p W$.   We first consider the case of
relatively weak bulk pinning when $I_p < (2/\pi)I_{s0}$, i.e., $p <  2/\pi.$  
In low fields ($0 < H_a < H_d$, region I of Fig.\ 1),   vortices nucleate on
the right-hand side at $x = W$ when
$I$ slightly exceeds
$I_s(H)$.  As long as
$K_z(x)=K_{Iz}(x)+K_{az}(x)$ exceeds  $K_p$, these vortices are driven
entirely across the strip, traveling rapidly (speed $v$
governed solely by the force-balance equation 
$[K_z(x)-K_p]
\phi_0 = \eta v d$, where $\eta$ is the viscous drag coefficient), and 
annihilating with their images on the opposite side of the strip.  The
critical current is then
$I_c(H_a,p) = I_s(H_a)$, and the normalized critical current is 
\begin{equation} i_c(h,p) = I_c(H_a,p)/I_{s0} = 1-h.
\label{icsmallh}
\end{equation} However, $K_{min}$, the minimum value of $K_z(x)$, decreases
with increasing $H_a$ and reaches
$K_p$ at
$I = I_s(H_a)$ when $H_a=H_d$, where
\begin{equation} h_d = H_d/(I_{s0}/2 \pi W) = \frac{1}{2} [1-(\pi p/2)^2].
\label{d1}
\end{equation} 
  
\begin{figure}
\epsfxsize=0.8 \hsize
\centerline{
\vbox{
\epsffile{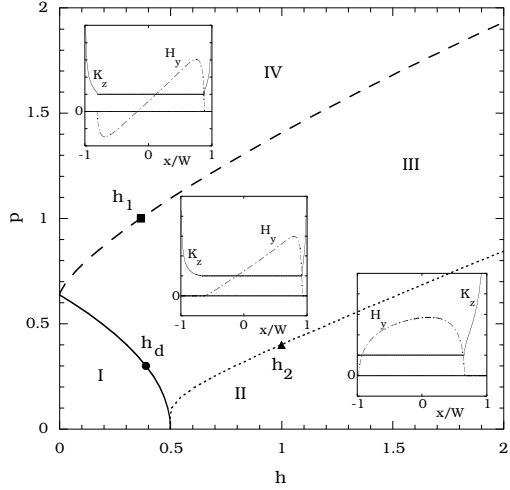} }}
\vskip \baselineskip
\caption{Behavior at the critical current vs reduced field $h$ and bulk
pinning parameter $p$.  In region I, the strip is vortex-free [$H_y(x,0)=0$]
and the sheet current density is everywhere above the bulk-pinning critical
value [$K_z(x)>K_p$]. In II, there is a vortex-free zone where $K_z(x)>K_p$ on
the right side of the strip and a vortex dome (dot-dashed curve in inset)
where 
$K_z(x)=K_p$ (solid curve in inset). In III, there are three zones: two
vortex-free zones on either side of a vortex dome.  In IV, there are four
zones: a vortex dome where $H_y(x,0)>0$, an antivortex dome where
$H_y(x,0)<0$, and two vortex-free zones where  $K_z(x)>K_p$ near the edges.
The curve $h_d(p)$ (solid) separates regions I and III, $h_1(p)$ (dashed)
separates III and IV, and 
$h_2(p)$ (dotted) separates II and III.}
\label{pvsh}
\end{figure}

When $p < 2/\pi$ and $h > h_d$ or when $p > 2/\pi$, i.e., for $h$ and $p$
outside  region I of Fig.\ 1, nucleated vortices stop inside the strip, and
dome-like flux distributions occur at  the critical current. To determine the
critical current with domes present, one must first calculate the vortex (and
antivortex) density
$n(x) = \mu_0 H_y(x,0)/\phi_0$ inside the dome, where $K_z(x)=K_p$, and the
sheet current density
$K_z(x)$ outside, where $n(x)=0$. We have obtained these mathematically by
using the Cauchy integral inversion
method\cite{MakELisJLet,MakELisAPL1,ILMakEPL,MakELisAPL2,Gakhov} to
invert the Biot-Savart law. However, we shall use
the method of complex fields\cite{MawaClem} to give a physical interpretation
of the mathematical results.

Following Ref.\cite{MawaClem}, we express the two-dimensional field
distribution as an analytic function $H(\zeta) = H_y(x,y) + i H_x(x,y)$ of the
complex variable $\zeta = x + i y$, such that the Biot-Savart law becomes
\begin{equation} H(\zeta) = H_a +
\frac{1}{2\pi}\int_{-W}^{W}\frac{K_z(u)}{\zeta-u}du.
\label{BScomplex}
\end{equation} The inversion procedure yields $H(\zeta)$ of the following form
at the critical current
$I_c(H_a)$:
\begin{equation} H(\zeta) =
\frac{(\zeta-a)^{1/2}(\zeta-b)^{1/2}}{(\zeta^2-W^2)^{1/2}}[H_a +
\frac{K_p}{2\pi}Q(a,b,\zeta)],
\label{HzetaI}
\end{equation} where
\begin{mathletters}
\begin{eqnarray} Q(a,b,\zeta) &=
\int_{a}^{b}\frac{\sqrt{W^2-u^2}}{(\zeta-u)\sqrt{(u-a)(b-u)}}du 
\label{Idef} \\ &=
\frac{2(W+a)}{\sqrt{(W-a)(W+b)}}[\frac{(W-\zeta)}{(\zeta-a)}
{\mathbf{\Pi}}(\frac{(\zeta+W)(b-a)}{(\zeta-a)(W+b)},q)\nonumber
\\   &   +{\mathbf{\Pi}}(\frac{b-a}{W+b},q)] ,
\label{IPis}
\end{eqnarray}
\end{mathletters} and 
\begin{equation} q^2 = \frac{2W(b-a)}{(W-a)(W+b)}.
\label{q}
\end{equation}  In Eq.\ (\ref{HzetaI}), the term proportional to $H_a$ is
simply the complex field describing the Meissner-state response to the applied
field $H_a$ of two parallel superconducting strips\cite{Benkr,AABB} ($-W < x <
a$ and $b < x < W$).  The term  proportional to $K_p$ is the complex field
describing the image-current response\cite{Norris72} of the two strips to
currents
$K_p du$ summed over the region $a < u < b$.  We have evaluated the integral
in Eq.\ (\ref{Idef}) in terms of complete elliptic integrals  of the third
kind\cite{Grad00,Abram67,Math,Selfridge58,Byrd}
${\mathbf{\Pi}}(n,k)$, where
$n$ is called either the characteristic or the parameter, and $k$ is called
the modulus.  
 Equation (\ref{IPis}) can be used to evaluate  $H_y(x,0) = {\rm Re} H(x)$ and
$K_z(x) = -2{\rm Im} H(x+i\epsilon)$.

For  $p > 2/\pi$ and small values of $h$, i.e., for $h$ and $p$ in region IV
of Fig.\ 1, the vortex distribution at the critical current can be described as
a double dome, consisting of a vortex dome adjacent to an antivortex dome (see
inset).  Just above the critical current, vortices nucleate at $x = W (x' =
1)$, where
$K_z(W-\Lambda) = K_s$,  move rapidly to the left through an otherwise
vortex-free region ($b < x < W$), and then move slowly to the left through a
vortex-filled region (the vortex dome), $a_+ < x < b$.   Antivortices nucleate
at $x = -W$, where $K_z(-W+\Lambda) = K_s$, move rapidly to the right
through an otherwise vortex-free region ($-W < x < a$), and then move slowly
to the right through an antivortex-filled region (the antivortex dome),
$a < x < b_-$.   Vortices and antivortices annihilate where the two domes meet
at $x = b_- = a_+$.

Two equations must be solved simultaneously for
$a$ and
$b$ at the critical current
$I_c(H_a)$ for known values of $h$ and $p$ in region IV of Fig.\ 1. One
condition is that
$K_z(W-\Lambda) = K_s$, which yields from Eqs.\ (\ref{HzetaI}) and (\ref{IPis})
\begin{eqnarray}
&\sqrt{(1-a')(1-b')}h+(1+a')\sqrt{\frac{1-b'}{1+b'}}{\mathbf{\Pi}}(\frac{b'-a'}{1+b'},q)p
\nonumber \\ &=1,
\label{K+W}
\end{eqnarray} where we use the normalized quantities $a'=a/W$ and $b' = b/W$.
 The other condition, that $K_z(-W+\Lambda) = K_s$, yields 
\begin{eqnarray}
&-\sqrt{(1+a')(1+b')}h+(1-b')\sqrt{\frac{1+a'}{1-a'}}{\mathbf{\Pi}}(\frac{b'-a'}{1-a'},q)p
\nonumber \\ &=1.
\label{K-W}
\end{eqnarray}

Expansion of Eq.\ (\ref{BScomplex}) for large $\zeta$ yields $H(\zeta) = H_a +
I/2 \pi \zeta +
\mathrm{O}(1/\zeta^2)$.  Expanding Eqs.\ (\ref{HzetaI}) and (\ref{Idef}) in
powers of
$\zeta$ and making use of Eq.\ (\ref{K+W}), we  obtain the normalized critical
current, $i_c = I_c/I_{s0}$:
\begin{eqnarray}  &i_c(h,p)  = -\frac{a'+b'}{2\sqrt{(1-a')(1-b')}} 
\nonumber \\  &+
\frac{1}{2}\sqrt{\frac{1+b'}{1-a'}} [(1-a'){\mathbf{E}}(q)
+(1+a'){\mathbf{K}}(q)]p,
\label{ic}
\end{eqnarray} where $a'$ and $b'$ are determined from Eqs.\ (\ref{K+W}) and 
(\ref{K-W}) for the desired values of $h$ and $p$.

The double-dome vortex-antivortex distribution (region IV of Fig.\ 1) occurs at
$i_c$ only for
$h$ in the range $0 < h < h_1$ for $p >2/\pi$.  Here, $h_1(p)$ is the lowest
value of $h$ that makes
$H_y(x,0)
> 0$ in the region
$a < x < b$. Thus, one of the equations determining $h_1$ is  $H_a + (K_p/2
\pi)Q(a,b,a+\epsilon) = 0$ [see Eq.\ (\ref{HzetaI})], which yields
\begin{eqnarray}  &h + 
\frac{1}{(b'-a')\sqrt{(1-a')(1+b')}}\nonumber \\ &[(1-a')(1+b'){\mathbf{E}}(q)
-(1+a')(1-b'){\mathbf{K}}(q)\nonumber
\\&-(b'-a')(1-b'){\mathbf{\Pi}}(\frac{b'-a'}{1-a'},q)]p = 0.
\label{H1}
\end{eqnarray}
$h_1$ can be determined for a given value of $p$  as the value of $h$ when
Eqs.\ (\ref{K+W}), (\ref{K-W}), and (\ref{H1}) are simultaneously solved for
$a', b',$ and $h$.

For known values of $h$ and $p$ in region III of Fig.\ 1, the left and right
boundaries of the vortex dome  $a'$ and $b'$ are determined by simultaneously
solving Eqs.\ (\ref{K+W}) and (\ref{H1}); Eq.\ (\ref{H1}) also gives the
condition that
$dK_z(x)/dx = 0$ at $x = a$. Once  $a'$ and $b'$ are found, Eq.\ (\ref{ic})
again can be used to calculate the critical current.  Just above the critical
current, vortices nucleate at
$x = W$, move rapidly to the left through the vortex-free region $b < x
< W$, move slowly to the left through the vortex dome
$a < x < b$, escape from the dome, and finally move rapidly to the left
through the vortex-free region
$-W < x < a$.

For increasing values of $h$, the left boundary of the vortex-filled region
moves closer to the left edge of the strip; $a$ becomes equal to $-W +
\Lambda$ when $h = h_2$, which can be determined for a given value of $p$ as
the value of $h$ when Eqs.\  (\ref{K+W}) and
 (\ref{H1}) are numerically solved for $b'$ and $h$,  taking $a' = -1 +
\Lambda/W$. 
(For Figs.\ 1 and 2, $\Lambda/W = 0.01$ was assumed.)
For $h>h_2$,  the field and current distributions and $i_c$ can  be
calculated with good accuracy by simply setting 
$a = W$ in Eq.\ (\ref{HzetaI}).  The complex field in region II of Fig.\ 1 is
then
\begin{equation}  H_{II}(\zeta) = \frac{(\zeta-b)^{1/2}}{(\zeta-W)^{1/2}}[H_a +
\frac{K_p}{2\pi}Q(b,\zeta)],
\label{HzetaII}
\end{equation} where
\begin{mathletters}
\begin{eqnarray} &Q(b,\zeta) =
\int_{a}^{b}\frac{\sqrt{W-u}}{(\zeta-u)\sqrt{b-u}}du 
\label{I2def} \\ &= 2 \sinh^{-1}{\sqrt{\frac{W+b}{W-b}}}  -2
\sinh^{-1}{\sqrt{\frac{(W+b)(\zeta-W)}{(W-b)(\zeta+W)}}}.
\label{Ilogs}
\end{eqnarray}
\end{mathletters}  
The condition
$K_z(W-\Lambda) = K_s$, which determines $b' = b/W$ at $I_c$ for $h$ and $p$
in region II of Fig.\ 1, becomes 
\begin{equation}
\sqrt{2(1-b')}[h+p \sinh^{-1}{\sqrt{(1+b')/(1-b')}}]=1
\label{K+WII}
\end{equation} instead of Eq.\ (\ref{K+W}), and the normalized critical current
can be expressed as 
\begin{equation} i_c(h,p)  = \frac{1}{4}\sqrt{2(1-b')} +
\frac{p}{2}\sqrt{2(1+b')}
\label{icII}
\end{equation} instead of Eq.\ (\ref{ic}).

\begin{figure}
\epsfxsize=0.8 \hsize
\centerline{
\vbox{
\epsffile{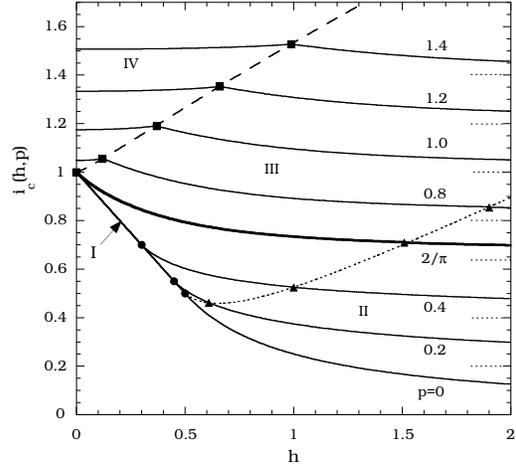} }}
\vskip \baselineskip
\caption{$i_c(h,p)$ (critical current normalized to $I_{s0}$) vs reduced field
$h$ for fixed values of the bulk pinning parameter $p$. The solid straight
line and solid circles denote values of
$i_c$ in region I, where
$h=h_d(p)$; the dashed curve and solid squares show $i_c$ at $h = h_1(p)$;
and  the dotted curve and solid triangles show $i_c$ at $h = h_2(p)$.   For
$p<2/\pi$, $i_c$ decreases linearly with
$h$ [Eq.\ (\ref{icsmallh})] up to $h_d$ and then decreases more slowly in
regions III [Eq.\ (\ref{ic})] and II [Eq.\ (\ref{icII})].  The bold curve shows
$i_c$ for the special case of $p = 2/\pi$.  For $p>2/\pi$,
$i_c$ increases by a few percent in the double-dome region IV and then
decreases more gradually in regions III and II.  In all cases, $i_c$
asymptotically approaches $p$ for large $h$ (short dotted lines along the
right side of the figure).}
\label{pvsh}
\end{figure} 

The reduced critical current $i_c$ as a function of $h$, calculated from Eq.\
(\ref{ic}) or (\ref{icII}), is shown in Fig.\ 2 for various values of $p$.  For
$p=0$ and $h>1/2$, we find $i_c = 1/4h$, as obtained in \cite {Kupr,Benkr} for
pin-free strips with an edge barrier.  In the opposite limit, $p \gg 1$, we
obtain $i_c \approx p$, as expected for bulk-pinning-dominated behavior.  For
$h \gg p$, we see from Eq.\ (\ref{K+WII}) that $b'$ approaches 1, and Eq.\
(\ref{icII}) yields $i_c \approx p + 1/4h.$  A generic feature of Fig.\ 2 is
the plateau in $i_c$ vs $h$ to the left of the dashed curve in region IV;
actually,
$i_c$ increases by a few percent as $h$ increases from 0 to $h_1$.  This
increase is due to a significant decrease in the width $b-a$ over which
$K_z(x)$ is restricted to $K_p$.  This effect is partially compensated by a
change in shape of $K_z(x)$ in the vortex-free zones [e.g., $dK_z(x)/dx = 0$ at
$x = a$ at $h = h_1$]. Field-dependent critical current densities $j_c(H_a)$
have been found experimentally in
Refs.\cite{Weyers,Muller,Aubin,Brion,Lefloch}, but the behavior was
interpreted solely in terms of  bulk pinning.

In agreement with earlier work,\cite {MakELisAPL2,Zhelezina} our results show
that the critical current of a strip is not a simple superposition of currents
$I_s$ and $I_p$, as was suggested in Refs.
\cite{Burlach,Tahara,Paltiel}. Only in the limit $h \gg p $ is it possible to
express the critical current as $I_c(H_a)=I_p+I_s(H_a)$.

We have assumed here that $K_p$ is a constant.  Because of the
nonlocal current-field relation [Eq.\ (\ref{BScomplex})], it would be a
challenging task to find  $I_c(H_a)$ when $K_p$ depends
upon the local magnetic field $H_y(x,0)$.

In summary, we have solved for the field dependence of the critical current
density in a superconducting strip accounting for both bulk pinning and a
geometrical edge barrier, and we have developed a procedure for finding the
magnetic-field and current-density distributions inside the strip at the
critical current.  In the presence of a strong edge barrier, we have found
strong field dependencies of the critical current.  Such effects should be
taken into account when interpreting experimental critical currents in low and
moderate magnetic fields. 

We thank Y. Mawatari for stimulating discussions.  This work was funded
by the Basic Research 
Foundation of Russia through Grant No. 01-02-16593, the Education Ministry
(Grant No. E00-3.4-331) and the Science  Ministry (Project   107-1(00)-P) of 
the Russian Federation, by Iowa State University of Science and Technology
under Contract No. W-7405-ENG-82 with the U.S. Department of Energy, and by the
National Academy of Sciences under the Collaboration in Basic Science and
Engineering Program supported by Contract No. INT-0002341 from the National
Science Foundation. \cite{boiler}

\end{multicols}
\end{document}